\documentclass[a4paper,twocolumn]{revtex4}  % journal-like
\usepackage[dvips]{graphicx}
\begin{document}

\title{Shot noise detection of the ultrasound tagged photon in
ultrasound-modulated optical imaging.}

\author{
M. Gross and P. Goy }

\affiliation{ Laboratoire Kastler Brossel
\thanks{Unite mixte de recherche du C.N.R.S. (UMR~8552),
  de l'Universite Pierre et Marie Curie et de l'Ecole Normale Supérieure
}\\
{ Departement de Physique de l'Ecole Normale Superieure }\\
{ 24, rue Lhomond  F-75231 Paris cedex 05,  France} }

\author{
M. Al-Koussa }

\affiliation{ Physical Department. Damascuss University.
Damascuss. Syria}

\date{\today}

\begin{abstract}
We propose a new detection method for ultrasound-modulated optical
tomography, which allows to perform, using a CCD camera,a parallel
speckle detection with an optimum shot noise sensitivity.
Moreover,we show that making use of a spatial filter system allows
us to fully filter off the speckle decorrelation noise. This
method being confirmed by a test experiment.
\end{abstract}


\maketitle

Ultrasound-modulated optical tomography is a new non invasive and
non ionizing biological tissues imaging technique. In this
technology an ultrasonic wave is focused into a diffusing medium
that scatters the incident optical beam. Due to the ultrasonic
vibration of the medium, some of the diffused photons are shifted
in frequency on an ultrasonic sideband. These are the so called
tagged photons \cite{leveque99}, which can be selected to perform
imaging. The advantage of the method is its combination of optical
contrast and ultrasonic resolution.

Many groups have worked on that field. Marks et al. \cite{Marks93}
investigated the modulation of light in homogeneous scattering
media with pulsed ultrasound. Wang et al. \cite{wang95} performed
ultrasound modulated optical tomography in scattering media. Lev
and al. made scattering media study in the reflection configuration
\cite{lev2000}. Wang and Ku \cite{wang98} developed a frequency
chirp technique to obtain scalable imaging resolution along the
ultrasonic axis by a 1D Fourier transform.  Leveque and al.
\cite{leveque99} performed parallel detection of multiple speckles
on a video camera and demonstrated an improvement of the detection
signal to noise ratio on 1D images of biological tissues. This
parallel speckle detection (PSD) is considered to be "so far, the
most efficient technique for ultrasound modulated optical
tomography"  \cite{wang2002}, and is now extensively used in the
field, combined \cite{leveque2000,wang2000b,forget2003} or not
\cite{wang2000,wang2002} with the frequency chirp technique.

By analysing the PSD detection process, we show nevertheless that
the PSD sensitivity is far from optimum. Moreover, as noticed by
Leveque et al. \cite{leveque99}, PSD is sensitive to the
"decorrelation of the speckle pattern, which reduces the signal
and increases the noise" (see also \cite{wang2000,wang2002}). In
this letter, we propose to solve these two problems by adapting
our heterodyne technique \cite{LeClerc2000} to PSD, i.e. by
performing heterodyne PSD (or HPSD). In HPSD, the  LO beam passes
outside the sample. The LO field is thus much larger, and the
detection sensitivity is much better. It is then possible to reach
the optimum shot noise limit. On the other side, the HPSD LO beam
is a plane wave. One can then separate the $k$-space components of
the detected field. By using a proper optical arrangement, which
reduces the $k$-space extend, we can fully filter off the decorrelation
noise.

Let us analyse the PSD detection process. The focus point of the
ultrasonic wave can be considered as a source of ultrasonic tagged
photons, which are detected coherently by heterodyne detection.
For a single pixel detector, the coherent detection selects the
field within the spatial mode that fits with the pixel considered
as an antenna. The collection efficiency can be characterized by
its "optical etendue", defined as the product of the emitting area
by the emission solid angle, which is the two-dimensional
generalization of the usual Lagrange invariant of geometrical
optics. Due to diffraction, for one single mode, the etendue is
about $\lambda^2$. For a $N$ pixels detector, since each pixel is
able to perform coherent detection within its mode, the etendue is
about $N \lambda ^2$. On the emission side, the etendue is about
$\pi S$, since each point of the sample external surface of area
$S$ diffuses photons in all outgoing directions (solid angle $\sim
\pi$). The collection efficiency $\eta$, which is the ratio of the
etendues is very low : $\eta \approx N\lambda ^2/\pi S$. For a $2
cm \times 2 cm \times 2 cm$ diffusing sample, $\eta \approx
10^{-10}$ for 1 pixel, and $\eta \approx 10^{-4}$ for $N=10^6$
pixels. The meaning of $\eta$ is quite simple. Forget et al.
\cite{forget2003} explained that, in order to detect the speckles
efficiently, it  is necessary to "position ... the camera ... to
match the size of a grain (of speckle) with the size of a pixel".
The camera must be thus placed quite far away from the sample, and
the photons, which are diffused by the sample in all directions,
have a probability $\sim \eta $ to reach the CCD.

On the CCD surface, each photon is converted into a photo electron
with a probability equal to the CCD quantum efficiency $Q$. In the
optimal case, the heterodyne detection noise is related to the
shot noise of the local oscillator. Accounting of the heterodyne
gain, this noise corresponds to 1 photo electron per pixel during
the measurement time \cite{bachor_1998} (since both shot noise and
heterodyne gain are proportional to the LO amplitude). PSD is far
from this optimum for two reasons. Firstly,  the PSD LO beam is
obtained by amplitude modulation of the main laser. The noise is
thus related to the total intensity (carrier + sideband), while
the gain depends on the sideband only. Secondly, the LO beam
passes through the sample and is diffused in many modes, which do
not match with the mode of the detector. Neglecting absorbtion,
and back reflection (which are also present), one looses here a
factor $\eta$ on the LO useful intensity. The LO intensity is then
too small to get enough heterodyne gain for efficient detection.

\begin{figure}[]
\begin{center}
\includegraphics[width = 8.5 cm,keepaspectratio=true]{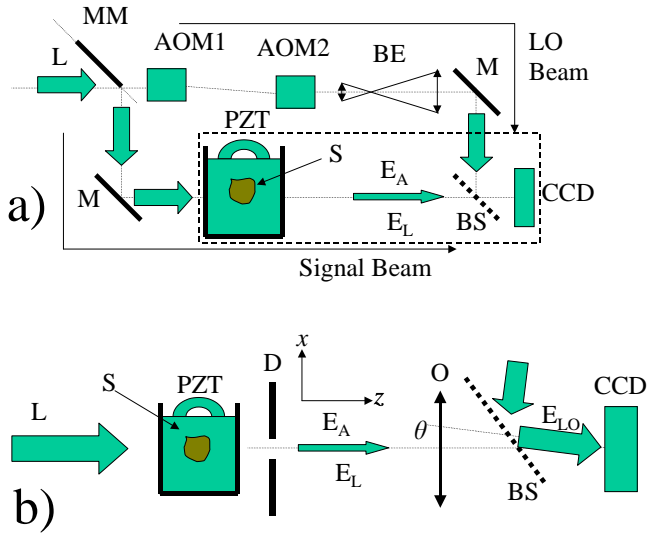}
\caption{a) Experimental setup. L: main laser, MM: moving mirror,
AOM1 et AOM2: acousto optic modulators, BE: beam expander, M:
mirror, PZT: ultrasonic transducer, S: sample, BS: beam splitter
and CCD: CCD camera. b) Spatial filter system.  $E_L$ : beam
diffused by the sample,  $E_A$: tagged photon beam, $E_{LO}$: LO
beam, S: sample, D: diaphragm with rectangular aperture, O: lens
and CCD: CCD camera. } \label{fig_setup}
\end{center}
\end{figure}

To improve the sensitivity, we propose to perform HPSD with the
Fig.\ref{fig_setup} a) setup. This setup is similar to the Toida
et al. \cite{inaba91} one, but with a CCD camera in place of the
mono pixel detector \cite{LeClerc2000}. The main laser L is a
$\lambda = 850 nm $, $20 mW$ Newport 2010M laser followed by an
optical isolator. The mirror MM splits the laser into two beams.
The low intensity LO beam is shifted in frequency by $\Delta f$ by
the 2 acousto optic modulators AOM1 ($80$ $MHz$) and AOM2 ($80$
$MHz$ $+ \Delta f$). It is expanded by BE ($20 \times $) in order
to get a plane wave (diameter $1.5$ $cm$) larger than the CCD
area. On the other side, the high intensity signal beam irradiates
the sample in a $13 cm$ width water vessel.

The PZT ultrasonic transducer (Parametrics: $f_a=2.2MHz$, diameter
$35 mm$, focal length: $50 mm$) generates an ultrasonic wave that
is focused into the sample. The signal beam that is diffused
toward the CCD ($z$ direction) at both the optical carrier
frequency (field $E_L$) and at the ultrasonic sidebands (field
$E_A$) interferes with the LO beam (beam splitter BS) on the CCD
camera. Accounting of the optical isolator, BS and water losses,
the measured laser power reaching the sample is $2.5$ $mW$. The
CCD camera (PCO Pixelfly: $1280 \times 1024$ pixels of $6.7 \times
6.7 \mu m $, $f_c=12.5$ $Hz$ , $12 bit$ digital, $2.2 \%$ measured
quantum efficiency at $850 nm$) records in real time the
interference pattern on a PC computer. MM is adjusted to get an
average of $2000$ shots per pixels ($1/2$ full scale) for the LO
beam, which remains ever much larger than the signal beam that is
strongly attenuated and diffused by the sample. To measure the
tagged photons field complex amplitude, we have chosen $\Delta f =
f_a + f_c/4$.

According to the camera gain given by PCO ($2.2$ electrons for 1
LSB: low significant bit), we have measured, without signal beam,
a noise corresponding  to 1 signal photo electron per pixel
(within $10 \%$). Our HPSD setup performs thus a optimal, shot
noise limited, heterodyne detection.

Consider now the speckle decorrelation noise. We have first to
notice that HPSD is less sensitive to this noise than PSD: the
noise is the same in both cases, but the signal is larger in HPSD,
because the heterodyne gain is higher. We will see now how to
filter off this noise. The setup (Fig.\ref{fig_setup}a dashed
rectangle) is modified as shown on Fig.\ref{fig_setup}b. The lens
$O$ (focal length $f=250$ $mm$) and the diaphragm $D$ ($\simeq 25
\times 5$ $mm$ located in the $O$ focal plane) collimate the field
diffused by the sample ($E_L$ and $E_A$), and reduce the $k$-space
extend in the $k_x$ direction. By this way, the speckle grains are
enlarged in $x$ and extended over several CCD pixels. Moreover,
the LO beam is slightly tilted (in the $x,z$ plane) making a angle
$\theta$ with the $z$ direction, so that the tagged photons versus
LO beam interference exhibit vertical fringes (along $y$). It is
then possible to separate the tagged photon signal (fringes) from
the decorrelation noise (no fringes) by a simple Fourier
calculation.

\begin{figure}[]
\begin{center}
\includegraphics[width = 8.5 cm,keepaspectratio=true]{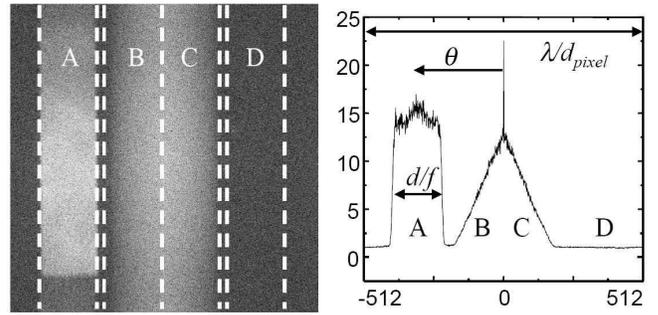}
\caption{Left hand side: $k$-space tagged photon intensity $\tilde
I(k_x,k_y)$. Right hand side: 1D plot of $\mathcal {I} (k_x)$.
Vertical scale is normalized with shot noise (noise in D zone). }
\label{fig_FFT}
\end{center}
\end{figure}

To illustrate this point, we have studied the field diffused by a
$3.5$ $cm$ phantom sample with a 15 Vpp (Volt peak to peak) $2
MHz$ excitation of the PZT. From $4$ successive CCD images signal
$I_1$,$I_2$...$I_4$ (where $I_i$ is proportional to $I=|E|^2$) we
have calculated the 4-phases complex signal $S=(I_1 - I_3) +
j.(I_2-I_4)$ where $j^2=-1$.  We have calculated $\tilde S
(k_x,k_y)=FFT\left( {S(x,y)} \right)$ by making a $1024 \times
1024$ truncation over the $S(x,y)$ data measured on the CCD ($1280
\times 1024$) followed by a Fast Fourier Transform (FFT). The
$\tilde I  (k_x,k_y)= |{\tilde S}|^2 $ matrix ($1024 \times 1024$)
is imaged on Fig.\ref{fig_FFT} (left hand side) with logarithmic
arbitrary gray scale. The sum over the column $\mathcal {I} (k_x)=
\sum\nolimits_{k_y} \tilde I((k_x,k_y))$ is also plotted  (right
hand side). One can separate here the contributions of the product
terms of $I=E.E^*$ (where $E=E_{LO}+E_{L}+E_{A}$: $E_{LO}$ being
the LO field, $E_{L}$ the field diffused at the carrier frequency
and $E_{A}$ the ultrasonic sideband tagged photons field). The
tagged photon heterodyne term $E_{LO}.{E_A}^*$ evolves fast is
space (fringes). It yields to the $A$ rectangular bright zone for
$\tilde I$, and to the $A$ peak for $\mathcal {I} $ (angular width
$d/f$, angular offset $\theta$). The speckle decorrelation noise
corresponds to the fluctuations on the $E_L.{E_L}^*$ term, which
evolves slowly in space (no fringes). It yields to the bright zone
in the center of the $k$-space image ($k_x \approx k_y \approx
0$), and to the triangular (convolution of 2 rectangles) peak in
the plot ($B$ and $C$). The fluctuations of the
$E_{LO}.{E_{LO}}^*$ term yield to the very narrow peak visible on
the 1D plot ($k_x=0$). The other terms give very small
contributions. For example, the $E_{LO}.{E_L}^*$ and
$E_{L}.{E_A}^*$ terms evolve fast in time ($2 MHz$) and are filter
off by the CCD. As seen, a proper choice of the $\theta$ tilt
allows us to separate in the $k$-space the tagged photon ($A$) and
the speckle decorrelation noise ($B$ and $C$) contributions to
signal. For control purpose, we have considered the zone $D$,
symmetric to $A$, where the shot noise only contributes.

\begin{figure}[]
\begin{center}
\includegraphics[width = 8.5 cm,keepaspectratio=true]{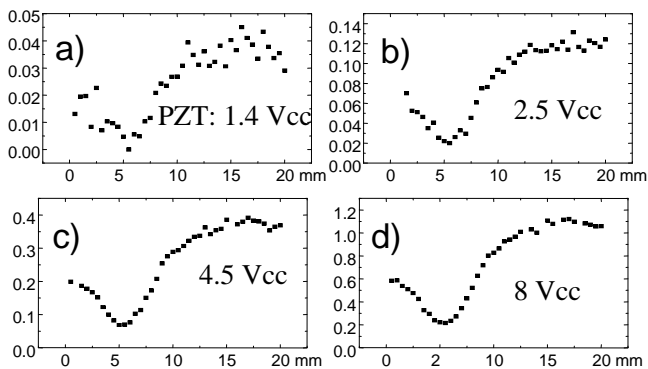}
\caption{$\mathcal {I}_{AD}/\mathcal {I}_{D}$ as a function of the
sample position for different ultrasonic excitation: $1.4$ (a),
$2.5$ (b) , $4.5$ (c), and $8$ (d) Vpp on the PZT transductor.
Laser equivalent power $2.5 mW$. CCD quantum efficiency $2.2\%$.
Measurement time $0.96 s$ per point. } \label{fig_data}
\end{center}
\end{figure}

We have performed a test experiment with the diffusing PSD sample
already studied in \cite{forget2003} (see Fig.2 of
\cite{forget2003}). In the sample, a $4 mm$ diameter vertical ($x$
direction) cylindric black inked zone absorbing the light.  The
sample is slightly compressed in the $z$ direction and its width
is $15 mm$. To get a pertinent information we have summed $\tilde
I (k_x,k_y)$ in the $A$ (tagged photons + shot noise) and $D$
(shot noise alone) zones: $\mathcal {I}_A = \sum\nolimits_{k_x \in
A} \mathcal {I}(k_x) $. We have calculated $\mathcal {I}_{AD} =
\mathcal {I}_{A}- \mathcal {I}_{D}$ (tagged photons alone), and
plotted $\mathcal {I}_{AD}/\mathcal {I}_{D}$ (tagged photon signal
normalized with respect to shot noise). The Fig.\ref{fig_data}a,
b, c and d show the plots obtained by moving the sample ($x=0...20
mm$ with $0.5 mm$ steps) with $1.4$, $2.5$, $4.5$, and $8$ $Vpp$
on the PZT respectively. Each point corresponds to $3$ successive
acquisitions of $4$ images ($0.96 s$). The black inked zone near
$x=5mm$ is clearly seen. As seen, outside the absorbing zone,
$\mathcal {I}_{AD}/\mathcal {I}_{D}$ ($\simeq 0.035$, $0.12$,
$0.37$ and $1.2$) is proportional to the square of the PZT applied
voltage (i.e. to the ultrasonic power). By making $\Delta f =
f_c/4$, with no ultrasonic wave, we have measured the carrier
field $E_L$ signal $\mathcal {I}_{AD}/\mathcal {I}_{D} \simeq
700$. The maximum ($8 Vpp$) ultrasonic conversion factor is then
$1.7 \times 10^{-3}$. As seen, curves c and d exhibit roughly the
same S/N (signal/noise) ratio. The technical noise is then the
limiting factor. On the other side, on curves a and b, the tagged
photon signal is lower and the S/N goes down. The shot noise
becomes thus the limiting factor. With $2$ zones ($A$ and $D$),
with $n_{pix} \approx 2.10^5$ pixels per zone, one expects for
$\mathcal {I}_{AD}/\mathcal {I}_{D}$ a shot noise of $\pm \sqrt
{2/n_{pix}}= \pm 0.003$, in good agreement with the noise observed
on curve a. This shows that the sensitivity obtained by our
technique is truly limited by shot noise.

As seen above, our HPSD detection method presents many advantages
for ultrasound-modulated tomography. It allows to perform parallel
speckle detection of the ultrasound-modulated component with an
optimum shot noise sensitivity, and to fully filter off the
speckle decorrelation noise. At the end, many controls are
possible on the data. One can measure , for example, both the
ultrasound-modulated signal (zone $A$), the shot noise (zone $D$)
and the speckle decorrelation noise (zone $B$ and $C$).

We thank the ESPCI group for their help (ultrasonic transducer,
diffusing sample), and for fruitful scientific discussions.

%\bibliographystyle{unsrt}

%\bibliography{diff}

\end{document}